\documentclass[runningheads]{llncs}
\usepackage{graphicx}
\usepackage{changepage}
\usepackage{wrapfig,lipsum,booktabs}

\begin{document}
\title{Blockchain-based Application Security Risks: \\
A Systematic Literature Review}
\titlerunning{Blockchain-based Application Security Risks}
%
\author{Mubashar Iqbal\inst{1} \and
Raimundas Matulevi\v{c}ius\inst{1}}
\authorrunning{M. Iqbal \& R. Matulevi\v{c}ius}
%
\institute{Institute of Computer Science, University of Tartu, Tartu, Estonia \\
\email{\{mubashar.iqbal,raimundas.matulevicius\}@ut.ee}}
\maketitle              
\begin{abstract}
Although the blockchain-based applications are considered to be less vulnerable due to the nature of the distributed ledger, they did not become the \textit{silver bullet} with respect to securing the information against different security risks. In this paper, we present a literature review on the security risks that can be mitigated by introducing the blockchain technology, and on the security risks that are identified in the blockchain-based applications. In addition, we highlight the application and technology domains where these security risks are observed. The results of this study could be seen as a preliminary checklist of security risks when implementing blockchain-based applications.

\end{abstract}
\section{Introduction}
Blockchain is a distributed immutable ledger technology \cite{Sato2018}. It gives participants an ability to share a ledger by peer-to-peer replication and updates every time when a transaction occurs. A ledger contains a certain and verifiable record of every single transaction ever made \cite{Lewis2015}. Security engineering is concerned with lowering the risk of intentional unauthorized harm to valuable assets to that level which is acceptable to the system’s stakeholders by preventing and reacting to malicious harm, misuse, threats, and security risks \cite{Firesmith2003}. Security plays an important role in blockchain-based applications. Those applications are acknowledged to be less vulnerable because the use of a decentralized consensus paradigm to validate the transactional information. They also backed by cryptography technology. However, the blockchain technology is continuously penetrating various fields and the involvement of the monetary assets raised the security concerns, mainly when the attackers stole the monetary assets or damage the system. For example, the reentrancy attack on the Ethereum based decentralized autonomous organization (DAO) smart contracts when an adversary gained control on \$60 million Ethers \cite{Atzei2017,Liu2018}.

Blockchain technology promises to overcome the security challenges, enhance the data integrity and to transform the transacting process into a decentralized, transparent and immutable manner. The recent progression of blockchain technology captured the interest of various sectors to transform their business processes by using blockchain-based applications. Hence, the security challenges are debatable and there is no comprehensive (or standardized) overview of security risks which can potentially damage the blockchain-based applications. There exist few studies reporting on security challenges in the blockchain platforms \cite{Atzei2017,Li2017}, but there is still a lack of focus on the blockchain-based application’s security.

In this paper, we present a systematic literature review (SLR) following the guidelines of \cite{Kitchenham2007}. Our research objectives are twofold. Firstly, we explain what security risks of centralized applications are mitigated by introducing blockchain-based applications. Secondly, we report the security risks of the blockchain-based applications which appear after introducing the blockchain technology. The main contributions of our study are: (1) a list of security risks in the blockchain-based applications which mitigate or inherit by incorporating the blockchain technology/platform, (2) aggregate a list of possible countermeasures and (3) an overview of the prominent research domains which are nourishing by the blockchain. The results of this study could be seen as a preliminary checklist of security risks when implementing blockchain-based applications.

The rest of the paper is structured as follows: Section 2 provides an overview of the blockchain and related work. Section 3 presents the contributions which explain the SLR process and Section 4 discuss its results. In Section 5, conclusion and future research directions are conferred.

\section{Background}
In this section, first, we introduce the blockchain technology. Second, we present an overview of related work.

\subsection{Overview of Blockchain Technology}
Blockchain forms a chain by a sequence of blocks that replicates over a peer-to-peer (P2P) network. In the blockchain, each block is attached to the previous block by a cryptographic hash, a block contains block header and a list of transactions as a Merkle tree. Blockchain is classified as a permissionless or permissioned \cite{Pradeepkumar2018}. In permissionless blockchain, anyone can join or leave the network and transactions are publicly available. In permissioned blockchain only predefined verified nodes can join the network and transactions visibility is restricted \cite{Ali2018,Pradeepkumar2018}.

In the blockchain, a smart contract (SC) is a computer program \cite{Atzei2017,Buterin2014} which constitutes a digital contract to store data and to execute functions \cite{Macrinici2018} when certain conditions are met. In the ethereum platform, developers use \emph{Solidity} programming language to write a smart contract and to build decentralized applications \cite{Buterin2014}. In Hyperledger Fabric, a smart contract is called chaincode. Similarly, other blockchain platforms introduce smart contracts to perform contractual agreements in a digital realm. The smart contracts are the high-level programming language-based programs and those can be error-prone where security flaws could be introduced (e.g. the reentrancy bug \cite{Liu2018}).

Blockchain eliminates the trusted intermediary and follows the decentralized consensus mechanism to validate the transactional information. Different blockchains use various consensus mechanism. Proof of Work (PoW) is a widely used computational rich energy-waste consensus strategy where special nodes called miners validate transactions by solving the crypto puzzle. Proof of Stake (PoS) is an energy-efficient consensus strategy \cite{Zheng2016} where miners become validators \cite{FabianVogelsteller2018} and lock a certain amount of cryptocurrency to show ownership to participate in the consensus process. There are other consensus mechanisms, for example, Delegated Proof of Stake (DPoS), Proof of Authority (PoA), Proof of Reputation (PoR) and Proof of Spacetime (PoSt).

The number of blockchain platforms is rapidly growing and thus, security becomes an important factor of the successful blockchain-based applications. In this paper, we focus on three frequently used blockchain platforms (Bitcoin, Ethereum, Hyperledger fabric). In addition, we also look at customised permissioned \& permissionless platforms (see Table~\ref{tab_stats}). Our goal is to learn which security risks and threats are considered in the applications of these platforms.

\subsection{Related Work}
There exist a few surveys, which consider blockchain platforms security risks. For instance, Li et al. \cite{Li2017} overview the security attacks on the blockchain platforms \& summarise the security enhancements. In our work, we consider the security risks on the \emph{blockchain-based applications} and their countermeasures.

Another related study \cite{Atzei2017} is conducted on Ethereum \emph{smart contracts} security. It reports on the major security attacks and presents a taxonomy of common programming pitfalls, which could result in different vulnerabilities. This study focuses on the security risks in the Ethereum smart contracts, further investigation is required to explore possible security risks in \emph{smart contracts based decentralized applications} and their viable countermeasures.

The main attributes of blockchain are integrity, reliability and security \cite{Koteska2017} which are also important in the IoT systems. The conventional approaches and reference frameworks of IoT network implementation are still unable to fulfil the requirements of security \cite{Khan2018}. Minhaj et al. \cite{Khan2018} survey major security issues of IoT and discuss different countermeasures along with the blockchain solution. This study, however, does not detail security challenges in the \emph{blockchain-based IoT applications}. Our study reviews the different blockchain-based IoT applications, discusses their security risks and potential countermeasures.

\section{Survey Settings}
In \cite{Kitchenham2007}, a comprehensive approach is presented to perform a SLR. In this section, we apply it to conduct a SLR on the security risks in the blockchain-based applications.

\subsection{Review Method}
In order to achieve the objectives of this study, we consider four research questions: (i) What are the domains where blockchain solutions are applied? (ii) What security risks are mitigated by the blockchain solutions? (iii) What do security risks appear within the blockchain-based applications? (iv) What are the countermeasures to mitigate security risks in the blockchain-based applications?

\textbf{Selection of databases}. The selection of electronic databases and literature search is carried out by consulting with the experts of software security. Literature studies are collected from ACM digital library, IEEE digital library, ScienceDirect, SpringerLink and Scopus. The \textbf{search queries} (including some alternative terms and synonyms) are formulated as follows: \\

\noindent \emph{Blockchain applications security (risks, threats, gaps, issues, challenges), permissioned blockchain applications security, permissionless blockchain applications security, public blockchain applications security} \\

\noindent \textbf{Relevance and Quality Assessment}. The inclusion and exclusion criteria listed in Table~\ref{tab_inclusion}. In this study, we only include the peer-reviewed literature because most of the grey literature is based on assumptions, abstract concepts and prejudices towards the security of their applications. Based on these shreds of evidence the grey literature could lead to the publication bias and erroneous results, so in order to eliminate these concerns only peer-reviewed literature is considered. 

\begin{table}
\renewcommand\arraystretch{1}
\caption{Inclusion and exclusion criteria.}\label{tab_inclusion}
\begin{tabular}{p{6cm}|p{6cm}}
\hline
Inclusion Criteria &  Exclusion Criteria\\
\hline
Only the peer-reviewed literature &  Literature that does not subject to peer review \\
\hline
Literature studies that discuss security risks in the blockchain-based applications &  Grey literature or informal studies with no concrete evidence \\
\hline
\end{tabular}
\end{table}

The selection of the studies was made after reading the paper \emph{title, abstract, introduction and conclusion} sections. Finally, following the quality guidelines of \cite{Kitchenham2007} and research scope of our study we have assessed the quality of studies using the following questions:

\begin{itemize}
    \item Are the goals and purpose of a study is clearly stated?
    \item Is the study describes security risks on the blockchain-based applications?
    \item Is the study provide the countermeasures to mitigate security risks?
    \item Is the study answered the defined research questions?
    \item How well the research results are presented?
\end{itemize}

\noindent The answers to the above questions are scored as follows: 1=Fully satisfy, 0.5=Partially satisfy, 0=Not satisfy. The studies with 2.5 or more points are included.

\subsection{Screening Results}

\begin{wraptable}{r}{4.5cm}
\begin{minipage}{1\linewidth}
\scriptsize
\renewcommand\arraystretch{1}
\vspace{-55pt}
\caption{Literature studies.}\label{tab_review_res}
\begin{tabular}{p{1.8cm}|p{0.8cm}|p{0.7cm}|p{0.7cm}}
\hline
 Database & Total & Excl. & Incl. \\
\hline
ACM & 21 & 11 & 10 \\
\hline
IEEE & 31 & 9 & 22 \\
\hline
ScienceDirect & 22 & 15 & 7 \\
\hline
SpringerLink & 23 & 12 & 11 \\
\hline
Scopus & 44 & 26 & 18 \\
\hline
Total & 141 & 73 & 68 \\
\hline
\end{tabular}
\end{minipage}%
\vspace{-30pt}
\end{wraptable}

Table~\ref{tab_review_res} presents the screening results. Initially, a total of 141 studies was collected. Later 73 studies were excluded by applying inclusion/exclusion and quality assessment criteria. Finally, 68 studies remained\footnote{Here is a list of these SLR studies: http://datadoi.ut.ee/handle/33/89}. The extracted information outlines the study identification, research problem, security risks and countermeasures. 

\section{Results and discussion}

\begin{wraptable}{r}{7.5cm}
\begin{minipage}{1\linewidth}
\scriptsize
\renewcommand\arraystretch{1}
\vspace{-34pt}
\caption{Statistics of literature studies as per year.}\label{tab_stats}
\begin{tabular}{p{0.7cm}|p{0.9cm}|p{1.2cm}|p{0.6cm}|p{0.8cm}|p{0.8cm}|p{0.95cm}|p{0.6cm}}
\hline
& \multicolumn{3}{|c}{Permissionless} & \multicolumn{2}{|c|}{Permissioned} & \multicolumn{2}{c}{} \\
\hline
 & Bitcoin & Ethereum & CPL & HLF & CP & Generic & Total\\
\hline
2016 & 2 & 0 & 0 & 0 & 0 & 0 & 2 \\
\hline
2017 & 7 & 3 & 8 & 1 & 2 & 1 & 22 \\
\hline
2018 & 9 & 15 & 3 & 8 & 8 & 1 & 44 \\
\hline
Total & 18 & 18 & 11 & 9 & 10 & 2 & 68 \\
\hline
\end{tabular}
\vspace{-10mm}
\end{minipage}%
\end{wraptable}

In this section, we present the SLR results. Table~\ref{tab_stats} shows how the field of blockchain-based applications is emerging every year. We observe that \emph{Ethereum-based} applications are gaining popularity among others. Also, permissioned blockchain platforms (\emph{Hyperledger Fabric (HLF)} \& \emph{Customised Permissioned (CP)}) are arising because of those support various industry-based use cases beyond cryptocurrencies. Practitioners also presented various \emph{Customised Permissionless (CPL)} platforms to achieve customised tasks and to overcome the limitations of other platforms. The term \emph{Generic} refers to studies where the blockchain type and platform is not mentioned.

\subsection{Applications Domains}
Table~\ref{tab_application_domains} presents the quantity of \emph{applications domains \& technology solutions} based on the different blockchain platforms. It shows \emph{Healthcare} is mostly
\begin{table}[!ht]
\vspace{-8pt}
\scriptsize
\renewcommand\arraystretch{1}
\caption{Research areas based on different blockchain platforms.}\label{tab_application_domains}
\begin{tabular}{p{3.4cm}|p{1.22cm}|p{1.22cm}|p{1.22cm}|p{1.22cm}|p{1.22cm}|p{1.22cm}|p{0.8cm}}
\hline
& \multicolumn{3}{|c}{Permissionless} & \multicolumn{2}{|c|}{Permissioned} & \multicolumn{2}{c}{} \\
\hline
 & Bitcoin & Ethereum & CPL & HLF & CP & Generic & Total\\
\hline
\multicolumn{8}{c}{Applications domains where blockchain is used.} \\
\hline
Healthcare & 0 & 3 & 1 & 2 & 4 & 1 & 11 \\
\hline
Resource monitoring \& Digital rights management & 1 & 3 & 2 & 0 & 2 & 1 & 9 \\
\hline
Financial & 2 & 1 & 1 & 1 & 0 & 0 & 5 \\
\hline
Smart vehicles & 1 & 0 & 1 & 1 & 2 & 0 & 5 \\
\hline
Voting & 1 & 1 & 0 & 2 & 0 & 0 & 4 \\
\hline
\multicolumn{8}{c}{Technology solutions where blockchain is used.} \\
\hline
Security layer & 6 & 7 & 1 & 0 & 1 & 0 & 15 \\
\hline
IoT & 2 & 2 & 1 & 2 & 2 & 0 & 9 \\
\hline
Total & 13 & 17 & 7 & 8 & 11 & 2 & 58 \\
\hline
\end{tabular}
\vspace{-8pt}
\end{table}
studied application domain and \emph{security layer} as a technology solution. Also, it indicates that Ethereum is widely used blockchain platform for building the decentralized applications.

\subsection{Security Risks}
Security risks result in harm to the system and its components \cite{Jouini2014}. In our study, the identified security risks are classified into two categories. (i) Security risks which are mitigated by introducing the blockchain-based applications (see Table~\ref{tab_app_risks}), and (ii) Security risks which appear within the blockchain-based applications (see Table~\ref{tab_bc_risks}). Table~\ref{tab_app_risks} presents the most common security risks which show that the researchers are utilizing the blockchain-based applications to overcome the limitations of centralized applications. For example, \emph{data tampering attack} is mitigated in \emph{Healthcare} applications and DDoS attack/Single point failure is resisted by decentralized distributed property of blockchain.

\begin{table}
\vspace{-8pt}
\scriptsize
\renewcommand\arraystretch{1}
\caption{Security risks which are mitigated by introducing blockchain applications.}\label{tab_app_risks}
\begin{tabular}{p{3.8cm}|p{1.1cm}|p{1.2cm}|p{1.1cm}|p{1.1cm}|p{1.1cm}|p{1.1cm}|p{0.8cm}}
\hline
& \multicolumn{3}{|c}{Permissionless} & \multicolumn{2}{|c|}{Permissioned} & \multicolumn{2}{c}{} \\
\hline
 & Bitcoin & Ethereum & CPL & HFL & CP & Generic & Total \\
\hline
Data tampering attack & 7 & 8 & 4 & 7 & 5 & 1 & 32 \\
\hline
DoS/DDoS attack & 7 & 7 & 5 & 3 & 2 & 1 & 25 \\
\hline
MitM attack & 3 & 6 & 2 & 2 & 0 & 1 & 14 \\
\hline
Identity theft/Hijacking & 1 & 0 & 3 & 0 & 0 & 1 & 5 \\
\hline
Spoofing attack & 2 & 0 & 1 & 0 & 1 & 0 & 4 \\
\hline
Other risks/threats & 6 & 4 & 2 & 1 & 2 & 2 & 17 \\
\hline
Total & 26 & 25 & 17 & 13 & 10 & 6 & 97 \\
\hline
\end{tabular}
\vspace{-5pt}
\end{table}

In addition to risks in Table~\ref{tab_app_risks}, other risks (found once or twice in the studies) are: Side-channel attack, Impersonation attack, Phishing attack, Password attack, Cache poisoning, Arbitrary attack, Dropping attack, Appending attack, Authentication attack, Signature forgery attack, Keyword guess attack, Chosen message attack, Audit server attack, Inference attack, Binding attack and Bleichenbach-style attack

Table~\ref{tab_bc_risks} represents the most common security risks which appear within the blockchain-based applications after introducing the blockchain technology. The table indicates the security risks, which have a high probability to make the blockchain-based applications vulnerable to attack. 

\begin{table}[!ht]
\vspace{-8pt}
\scriptsize
\renewcommand\arraystretch{1}
\caption{Security risks which appear within the blockchain applications.}\label{tab_bc_risks}
\begin{tabular}{p{3.8cm}|p{1.1cm}|p{1.2cm}|p{1.1cm}|p{1.1cm}|p{1.1cm}|p{1.1cm}|p{0.8cm}}
\hline
& \multicolumn{3}{|c}{Permissionless} & \multicolumn{2}{|c|}{Permissioned} & \multicolumn{2}{c}{} \\
\hline
 & Bitcoin & Ethereum & CPL & HLF & CP & Generic & Total\\
\hline
Sybil attack & 5 & 1 & 1 & 4 & 1 & 1 & 13 \\
\hline
Double spending attack & 4 & 1 & 2 & 2 & 0 & 1 & 10 \\
\hline
51\% attack & 3 & 3 & 1 & 0 & 0 & 1 & 8 \\
\hline
Deanonymization attack & 2 & 1 & 3 & 0 & 0 & 1 & 7 \\
\hline
Replay attack & 2 & 4 & 1 & 0 & 0 & 0 & 7 \\
\hline
Quantum computing threat & 0 & 1 & 1 & 2 & 0 & 1 & 5 \\
\hline
Selfish mining attack & 1 & 0 & 2 & 1 & 0 & 0 & 4 \\
\hline
SC reentrancy attack & 0 & 2 & 0 & 0 & 0 & 1 & 3 \\
\hline
Other risks/threats & 6 & 1 & 6 & 3 & 1 & 3 & 20 \\
\hline
Total & 23 & 14 & 17 & 12 & 2 & 9 & 77 \\
\hline
\end{tabular}
\vspace{-5pt}
\end{table}
Hence the \emph{Sybil attack}, \emph{Double spending attack} and \emph{51\% attack} are the most appeared security risks after incorporating the blockchain technology. Other security risks which are appeared once or twice in the studies are: Eclipse attack, BWH attack, 25\% attack, Stake grinding attack, Block Discarding attack, Difficulty Raising attack, Pool-hopping attack, Node masquerading attack, Timestamp attack, Balance attack, Signature forgery attack, Confidentiality attack, Private keys compromise, Overspending attack, Collusion attack and Illegal activities.

In Table~\ref{tab_domains_risks} we encompass the security risks along with the blockchain-based applications research areas to show which security risks are more frequently occurring on different blockchain-based applications. Most frequently the security risks expose in \emph{Resource monitoring and digital rights management} applications, followed by the \emph{Financial}, \emph{Healthcare}, \emph{Smart vehicles} and \emph{Voting} applications. Also, blockchain is presented as a technology solution where researchers incorporated the blockchain as a security layer to protect against the listed security risks. However, Table~\ref{tab_domains_risks} shows 34 different security risks \emph{(combining both security risks which are mitigated and appear by introducing the blockchain solution)}. Furthermore, a blockchain technology solution for IoT based applications is rapidly increasing because it provides integrity, reliability and security \cite{Khan2018} and these are important for IoT based solutions to reach high requirements of security. By the results, the most common security risks in IoT based applications are mitigated by implementing the blockchain-based solution and only 3 different security risks are inherited after introducing the blockchain solution. The \emph{other} column represents the generic blockchain-based applications and blockchain technology solutions where no specific domain is studied.

\begin{table}[!ht]
\vspace{-8pt}
\scriptsize
\renewcommand\arraystretch{1}
\caption{Security risks based on the research areas.}\label{tab_domains_risks}
\begin{tabular}{p{3.15cm}|p{1.3cm}|p{1.1cm}|p{1.15cm}|p{0.95cm}|p{0.83cm}|p{1cm}|p{0.45cm}|p{0.6cm}|p{0.6cm}}
\hline
\multicolumn{8}{c}{Security risks which are mitigated by introducing blockchain applications.} \\
\hline
& \multicolumn{5}{c}{Applications} &  \multicolumn{2}{|c|}{Technology} & \\
\hline
& Healthcare & Resource monit. & Financial & Smart vehicles & Voting & Security layer & IoT & other & Total\\
\hline
Data tampering attack & 6 & 5 & 1 & 4 & 3 & 2 & 5 & 6 & 32 \\
\hline
DoS/DDoS attack & 0 & 5 & 1 & 3 & 1 & 7 & 3 & 5 & 25\\
\hline
MitM attack & 1 & 4 & 1 & 1 & 1 & 2 & 2 & 2 & 14 \\
\hline
Identity theft/Hijacking & 1 & 2 & 0 & 0 & 0 & 0 & 1 & 1 & 5 \\
\hline
Spoofing attack & 0 & 0 & 0 & 0 & 1 & 0 & 1 & 2 & 4 \\
\hline
Other risks/threats & 2 & 0 & 1 & 0 & 1 & 5 & 5 & 3 & 17 \\
\hline
\multicolumn{8}{c}{Security risks which appear within the blockchain applications.} \\
\hline
Sybil attack & 1 & 1 & 1 & 1 & 2 & 1 & 1 & 5 & 13 \\
\hline
Double spending attack & 0 & 4 & 2 & 0 & 0 & 2 & 0 & 2 & 10 \\
\hline
51\% attack & 0 & 4 & 0 & 0 & 1 & 1 & 0 & 2 & 8 \\
\hline
Deanonymization attack & 0 & 2 & 1 & 1 & 1 & 1 & 1 & 0 & 7\\
\hline
Replay attack & 0 & 2 & 1 & 0 & 0 & 4 & 0 & 0 & 7 \\
\hline
Quantum comp. threat & 1 & 0 & 0 & 0 & 0 & 2 & 0 & 2 & 5 \\
\hline
Selfish mining attack & 0 & 1 & 1 & 0 & 0 & 2 & 0 & 0 & 4 \\
\hline
SC reentrancy attack & 0 & 0 & 0 & 0 & 0 & 3 & 0 & 0 & 3 \\
\hline
Other risks/threats & 0 & 11 & 5 & 0 & 0 & 2 & 1 & 1 & 20 \\
\hline
Total & 12 & 41 & 15 & 10 & 11 & 34 & 20 & 31 & 174 \\
\hline
\end{tabular}
\vspace{-22pt}
\end{table}

\subsection{Countermeasures}
In this section, we overview countermeasures to mitigate the security risks listed in Table~\ref{tab_app_risks} and~\ref{tab_bc_risks}.

\textbf{Countermeasures introduced with blockchain solution}. The security risks presented in Table~\ref{tab_app_risks} are mitigated by implementing the blockchain-based applications together with the techniques to mitigate these risks. For instance, \emph{Data tampering attack} poses a threat to data-sensitive applications. In \cite{Yu2018,Zhang2018} authors implement the smart contract to mitigate votes tampering. In \cite{Sylim2018,Yu2018} authors encrypt information and associate a unique hash. Lei et al. \cite{Chen} propose a random oracle model with strong RSA. And Li et al. \cite{Li2018} introduce an elliptic curve digital signature algorithm (ECDSA) based signature scheme for anonymous data transmission along Merkle hash tree based selective disclosure mechanism. Han et al. \cite{Han2018} propose to use permissioned blockchain where only the authorized nodes are able to access the data as well as generate a cypher-text by using digital signatures.

\emph{DoS/DDoS attack} is another exploitable cyber-attack, it is resisted by a distribution of service on different nodes \cite{Yu2018}. The \cite{Lin2018,Decusatis2018} authors implement an access control scheme to prevent unauthorized requests. Androulaki et al. \cite{Androulaki2018} propose a block-list to track suspicious requesting nodes and the authors of \cite{Androulaki2018,Qin2018} incorporate the transaction fee to resist it. In order to resist the \emph{MitM attack}, authors suggest to encrypt an information \cite{Dagher2018,Yu2018} and publish on the blockchain \cite{Yu2018}. In \cite{Lin2018,Yao2018} research studies, an authentication scheme is introduced to verify each communication node. \emph{Identity theft/Hijacking} based risks are mitigated by information authentication and message generation time-stamping \cite{Fan2018}. Mylrea et al. \cite{Mylrea2017} suggest a permission-based solutions (e.g. KSI). \emph{Spoofing attack} is mitigated by introducing an anonymous communication among nodes \cite{Cebe2018} and Keyless Signature Infrastructure (KSI) based distributed \& witnesses trust anchor \cite{Mylrea2017}.

\textbf{Countermeasures to mitigate security risks of blockchain solutions}. The blockchain solution comes with a few trade-offs and inherits several security risks (see Table~\ref{tab_bc_risks}) of blockchain technology which are mitigated by implementing the various techniques, those techniques are listed below as countermeasures. In order to mitigate the \emph{Sybil attack}, in \cite{Gallo2018,Zhang2018} authors suggest the permissioned blockchain-based application. Bartolucci et al. \cite{Bartolucci2018} incorporate the transaction fee \& identification system to allow only authorized users to perform different operations. In \cite{Qin2018}, authors use the PoR scheme and Liu et al. \cite{Liu2018a} implement the customised blockchain to control the computing power. \emph{Double spending attack} is mitigated by the transaction verification based on unspent transaction state \cite{Androulaki2018}. In \cite{Alcarria2018} authors resisted this attack by PoA scheme and in \cite{Buchmann2017} by PoW complexity. Also, the Muzammal et al. \cite{Muzammal2018} append the nonce with each transaction. Another frequent security risk on the blockchain-based applications is \emph{51\% attack} which is resisted by implementing trusted authorities control \cite{Zhu2018} and Hjalmarsson et al. \cite{Hjalmarsson2018} customised the Ethereum blockchain to permissioned blockchain. 

In order to mitigate \emph{Deanonymization attack}, in \cite{Lin2018} authors propose a solution to obtain identity information only after authorization. Bartolucci et al. \cite{Bartolucci2018} propose the mixer for mixing the position of output addresses. In \cite{Saritekin2018,Tosh2017} authors propose another solution to mitigate this attack by using the fresh key for each transaction. \emph{Selfish mining attack} is mitigated by PoR scheme \cite{Qin2018} and by raising the threshold \cite{Tosh2017}. No countermeasure is found for \emph{Replay attack}. In order to overcome the \emph{Quantum computing threats}, Yin et al. \cite{Yin2018} implement the lattice cryptography and in \cite{Buchmann2017} authors suggest an additional digital signature or a hard fork in the post-quantum era. Decusatis et al.  \cite{Decusatis2018} propose a need of quantum blockchain. To eliminate the chances of \emph{Smart contract reentrancy attack}, authors of \cite{Liu2018} present the automation tool to detect smart contract bugs via run-time trace analysis and in \cite{Tikhomirov2018} authors built a static analysis tool that detects reentrancy bugs in a smart contract and translates solidity source code into an XML-based intermediate representation and checks it against XPath patterns.

\section{Conclusion and Future Work}
In this paper, we present a systematic literature review on the blockchain-based applications security risks to explain what security risks are mitigated by introducing the blockchain-based applications, and what security risks are reported in the blockchain-based applications. Our result is a preliminary checklist to support developers' decisions while developing blockchain-based applications.

Our current study has a few limitations: (i) Applications which are built on the blockchain platforms are mostly in the prototype phase. Thus the research studies present only the conceptual illustrations of different security risks and their countermeasures but not the real-life applications. (ii) The field of decentralized applications is relatively new but continuously evolving. Not all the possible security risks are researched in the blockchain-based applications which show the possibility that a wide range of security risks will emerge in upcoming years. (iii) This study found that a lot of security risks and their countermeasures are either obscure or the practical implementation is still not available. Overcoming these limitations could possibly result in the interesting insights and contribute to the explaining the blockchain-based application security risks, their vulnerabilities and the countermeasures for more in-depth.

As a part of the future work, our aim is to build a comprehensive reference model for security risk management to systematically evaluate the security needs. This model would explain the protected assets of the blockchain-based applications, and countermeasures to mitigate their risks. \\

\textbf{Acknowledgement}. This research has been supported by the Estonian Research Council (grant IUT20-55). 

%
%
%
%
\bibliographystyle{splncs04}
\bibliography{mybibliography}

\end{document}